\documentclass[aps,twocolumn,showpacs]{revtex4}
\usepackage{graphicx}

\newcommand{\ket}[1] {\left| #1 \right\rangle}

\begin{document}

\title{Using Superconducting Qubit Circuits to Engineer Exotic Lattice Systems}

\author{Dimitris I. Tsomokos$^1$, Sahel Ashhab$^{2,3}$, Franco Nori$^{2,3}$}
\address{$^{1}$Department of Mathematics, Royal Holloway, University of London, Egham TW20 0EX, UK \\
$^{2}$Advanced Science Institute, The Institute of Physical and Chemical Research (RIKEN), Wako-shi, Saitama 351-0198, Japan \\
$^{3}$Physics Department, University of Michigan, Ann Arbor,Michigan 48109-1040, USA}

\date{\today}

\begin{abstract}
We propose an architecture based on superconducting qubits and
resonators for the implementation of a variety of exotic lattice
systems, such as spin and Hubbard models in higher or fractal
dimensions and higher-genus topologies. Spin systems are realized
naturally using qubits, while superconducting resonators can be
used for the realization of Bose-Hubbard models. Fundamental
requirements for these designs, such as controllable interactions
between arbitrary qubit pairs, have recently been implemented in
the laboratory, rendering our proposals feasible with current
technology.
\end{abstract}

 \maketitle


\section{Introduction}

The idea of quantum simulation has led to a large number of
theoretical proposals and some remarkable experimental results
over the past decade \cite{Buluta}. For example, the transition
between a superfluid and a Mott insulator, which is a classic
condensed-matter-physics paradigm, has been investigated in a
controlled fashion using a gas of ultracold atoms \cite{Greiner}.
Based on the success of such experiments, there have been intense
efforts to devise methods for the simulation of various physical
problems whose theoretical analysis is challenging. One area that
has been studied substantially in the theoretical literature, with
a number of fundamental questions still unanswered, is that of
lattice systems in arbitrary dimensions and topologies.
Implementing such systems using naturally occurring systems in two
or three dimensions is challenging because the required
connectivity is incompatible with the geometry of the \emph{physical} space in
which the spins or lattice sites reside. Here we propose the
implementation of such exotic systems using electronic nanocircuits
based on superconducting qubits (SQs) and superconducting resonators \cite{You}.

The basic elements required for the implementation of our
proposals have all been demonstrated experimentally. In
particular, in a recent experiment \cite{Harris}, Harris \emph{et
al.} fabricated a SQ circuit with $N>100$ qubits and demonstrated
fully controllable interactions in blocks of 8 qubits,
i.e.~essentially \emph{all pairs} of qubits were coupled to each
other with individually controllable coupling strengths. This
ability to design at will tunable couplings between any pair of
qubits is a crucial ingredient in our proposal for experimentally
engineering exotic quantum architectures, which are hard to study
otherwise. We investigate a range of systems that can be
implemented using such SQ circuits with fully controllable
interactions. The experiments that we propose should be realizable
in the near future, particularly those related to the
thermodynamic properties of classical spin systems, where
many-qubit coherence is not required.

We start by reviewing the present-day technology of
superconducting qubits, resonators and tunable couplers. We then
go through our list of proposals. In particular, we examine
complex quantum systems in \emph{higher} and
\emph{nonconventional} dimensions, such as spin lattices in
noninteger (fractal) dimensions
\cite{Fractals_book,Ising_Fractal_Dim,Tissier}, and we consider
the embedding of lattice systems on \emph{exotic topologies}, such
as the Klein bottle and M\"{o}bius strip
\cite{Ising_ExoticTopo,Mobius_molecule,Topo_review}. Both types of
systems (i.e.~those involving nonconventional dimensions or
topologies) could advance our understanding of phase transitions
\cite{Wu_Book,Sachdev,Fazio_review} and lead to practical
applications. For example, it is known that dimensionality plays
an important role in phase transitions \cite{Wu_Book}, and the
ability to engineer any desired dimension would provide a valuable
experimental knob in their study. Furthermore, spin lattices in
nontrivial topologies and in higher dimensions have been shown to
be better suited for implementing passive
quantum-information-protecting schemes than conventional systems
\cite{Dennis}. Near the end of the paper, we discuss the
implications of the \emph{in-situ} tunability of the parameters
and the possibility of performing novel \emph{topology-quench}
experiments.

\section{Physical Implementation}

\begin{figure}
  \centering
  \resizebox{0.8\linewidth}{!}{\includegraphics{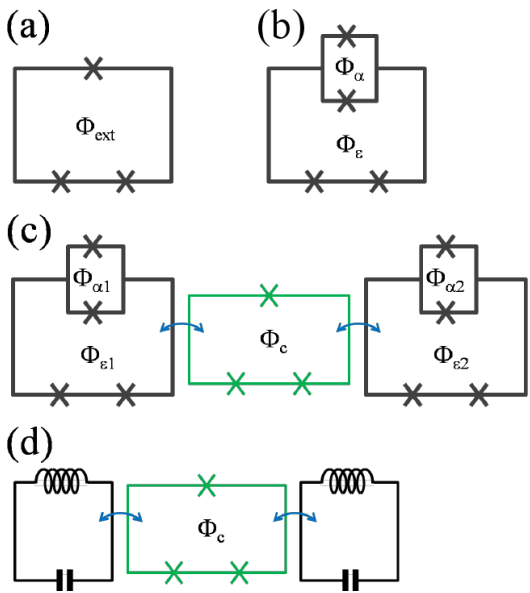}}
  \caption{(Color online) Schematic diagrams of (a) a simple three-junction flux qubit, (b) a flux qubit with a tunable gap, (c) two flux qubits interacting through a tunable coupler, and (d) two resonators interacting through a tunable coupler. In (a) the external magnetic flux $\Phi_{\rm ext}$ threading the superconducting loop controls the parameter $\epsilon$ in the single-qubit Hamiltonian (Eq.~1). In (b) the magnetic flux $\Phi_{\epsilon}$ controls the parameter $\epsilon$ in the Hamiltonian, while $\Phi_{\alpha}$ controls the parameter $\Delta$. In (c) two flux qubits are coupled inductively to a common coupler, resulting in an effective coupling between the two qubits. The effective coupling strength $J$ can be tuned through the magnetic flux $\Phi_{\rm C}$. In (d) two LC circuits, i.e.~resonators, are effectively coupled to each other through a tunable coupler.} \label{fig_1}
\end{figure}

Although various types of SQs could be used to implement the
architectures that we propose here, we shall focus on flux qubits,
since these were used in the experiments of Ref.~\cite{Harris}.
The flux qubit is a superconducting loop interrupted by a number
of Josephson junctions, typically one to four \cite{You}, as
illustrated in Fig.~\ref{fig_1}(a). When a magnetic flux close to
half a flux quantum is applied through the loop, two quantum
states (one with clockwise and the other with counterclockwise
circulating current) are almost degenerate and form the qubit
basis. These states are usually denoted by $\ket{\uparrow}$ and
$\ket{\downarrow}$, and the qubit can be thought of as a
spin-$1/2$ particle. The single-qubit Hamiltonian in the language
of Pauli matrices $\hat{\sigma}^{x,y,z}$ is
\begin{equation}
\hat{H}_{\rm q} = \frac{\Delta}{2} \hat{\sigma}^x +
\frac{\epsilon}{2} \hat{\sigma}^z,
\end{equation}
where $\Delta$ is the minimum gap at the half-flux bias point and
$\epsilon$ is the deviation from this point. It is worth noting
here that $\epsilon$ is an easily tunable parameter and that it
has recently become possible to also tune $\Delta$ using the
externally applied magnetic fields \cite{Paauw}, as shown in
Fig.~\ref{fig_1}(b).

If two qubits are placed next to each other, the magnetic-dipole
interaction gives rise to a two-qubit coupling term of the form
\begin{equation}
\hat{H}_{\rm int,q-q} = J \hat{\sigma}_1^z \hat{\sigma}_2^z,
\end{equation}
where $J$ is the coupling strength and the subscripts indicate the
different qubits. For two directly coupled qubits, the coupling
strength $J$ is fixed by geometry and material properties. One
could avoid this limitation and obtain an effectively tunable
coupling strength by employing a coupler, an additional circuit
element that mediates coupling between the two qubits
\cite{Averin}. A schematic diagram of this technique is shown in
Fig.~\ref{fig_1}(c). By tuning the bias parameters (e.g.~the
magnetic flux through the coupler's loop), one can effectively
tune the inter-qubit coupling strength $J$. An additional
advantage of using couplers is the flexibility allowed in their
design, which leads to the ability to couple qubits that are
separated by large distances and to produce coupling terms in any
desired pairing of the qubits \cite{Harris}. With the above
architecture, one obtains the many-qubit Hamiltonian
\begin{equation}
\hat{H_1} = \sum_i \frac{1}{2} \left( \Delta_i \hat{\sigma}_i^x +
\epsilon_i \hat{\sigma}_i^z \right) + \sum_{i,j} J_{i,j}
\hat{\sigma}_i^z \hat{\sigma}_j^z,
\end{equation}
with at least the parameters $\epsilon_i$ and $J_{i,j}$ being
tunable \emph{in situ}. Further details concerning the circuit can
be found in Ref.~\cite{Harris}.

While qubits are suited for the implementation of spin-lattice
Hamiltonians, Bose-Hubbard Hamiltonians require
harmonic-oscillator-like circuit elements, i.e.~resonators. These
can be implemented either as lumped-element LC circuits, as
illustrated in Fig.~\ref{fig_1}(d), or as coplanar-waveguide
resonators \cite{You}. In both cases, the resonator behaves as a
linear oscillator with the Hamiltonian
\begin{equation}
\hat{H}_{\rm osc} = \hbar\omega \left( \hat{a}^{\dagger} \hat{a} +
\frac{1}{2} \right),
\end{equation}
where $\omega$ is the oscillator frequency and $\hat{a}^{\dagger}$
and $\hat{a}$ are the oscillator's creation and annihilation
operators, respectively. Several experiments have demonstrated
coherent coupling between resonators and qubits \cite{You}. With
the technology that has been developed in that context, there
should be no difficulty in coupling resonators to each other using
couplers, thus leading to tunable coupling with any desired
pairing of the resonators. The interaction Hamiltonian is then
given by
\begin{equation}
\hat{H}_{\rm int,osc-osc} = J \left( \hat{a}^{\dagger}_i \hat{a}_j
+\hat{a}^{\dagger}_j \hat{a}_i \right),
\end{equation}
where the subscripts
indicate the different resonators. In writing this form for the
Hamiltonian, we have assumed that $J \ll \hbar\omega/n$ (with $n$
being the typical number of excitations in each resonator) such
that the rotating-wave approximation is valid. The resonators can
now be seen as sites in a Hubbard-like model, and the number of
excitations in any given resonator represents the number of
(bosonic) particles occupying that site. One thus obtains the
non-interacting Bose-Hubbard Hamiltonian
\begin{eqnarray}
\hat{H_2} =  \sum_i \hbar \omega_i \left( \hat{a}^{\dagger}_i
\hat{a}_i + \frac{1}{2} \right) + \sum_{i,j} J_{i,j} (
\hat{a}^{\dagger}_i \hat{a}_j +\hat{a}^{\dagger}_j \hat{a}_i ),
\end{eqnarray}
with the parameters $J_{i,j}$ being tunable \emph{in situ}. The
resonator frequencies $\omega_i$ generally exhibit small
deviations from the values specified when designing the circuit,
and recent experiments have demonstrated resonators with tunable
values of $\omega$ \cite{Sandberg}.

\section{Nonconventional Dimensions}

On the theoretical side, spin lattice systems in higher dimensions
($d \ge 3$) have been studied extensively in the past
\cite{Wu_Book}. Interest in these higher-dimensional systems stems
from the important role that dimensionality plays in the physics
of phase transitions and critical phenomena, as well as in
determining the magnetic and thermodynamic properties of
materials. Over the years, a number of different theoretical
methods have been used in studying higher-dimensional systems,
including renormalization-group and Monte-Carlo methods. However,
the validity of these methods generally cannot be established
rigorously \cite{Lundow}. Therefore, experimental investigation of
higher-dimensional systems is highly desirable. The experimental
realization of such systems is hindered, however, by the
difficulties associated with coupling spatially separated
elements, such as distinct two-level systems.

\begin{figure}
  \centering
  \resizebox{0.85\linewidth}{!}{\includegraphics{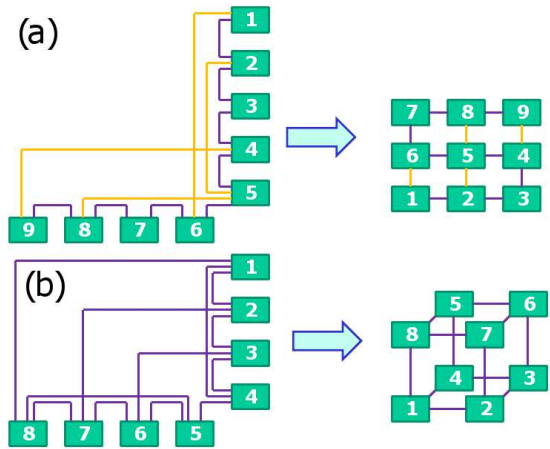}}
  \caption{(Color online) Engineering the effective dimensionality of a lattice system by changing the connections between qubits: (a) Nine qubits can be connected so as to form a linear chain [purple (black) connections] or a $3\times 3$ square lattice [purple (black) and orange (gray) connections]. (b) Eight qubits can be connected into a three-dimensional cube.} \label{fig_2}
\end{figure}

With SQ networks it should be possible to overcome the difficulty
of arbitrary connectivity \cite{Strauch}. The Ising model in $d =
1,2,3, \ldots$ dimensions can be implemented using the connections
illustrated in Fig.~\ref{fig_2}. In this context one might worry
whether the crossing of some of the lines in Fig.~2 would be a
problem. However, this is not the case, since such lines can be
fabricated in different layers, similarly to what was done in
Ref.~\cite{Harris}. Naturally there would be physical limits on
the number of layers in a realistic system, and one might worry
that for larger numbers of qubits a larger number of layers will
be required. However, the number of layers does not depend on the
system size, but only on the engineered effective dimension: in
principle no overlapping connections are needed for $d=2$, only
two layers of couplers are needed for $d=3$, and so on.

Similarly to higher-dimensional systems, spin systems with
noninteger dimensions have also received much theoretical interest
over the past few decades but have not been implemented
experimentally \cite{Fractals_book,Ising_Fractal_Dim,Tissier}. One
example of interesting theoretical results in this context is that
the Ising model can exhibit spontaneous magnetization at finite
temperatures on spin lattices with $d < 2$
\cite{Ising_Fractal_Dim}; another is the prediction that certain
spin models change critical behavior from second to first order at
specific noninteger dimensions \cite{Tissier}.

\begin{figure}
  \centering
  \resizebox{0.9\linewidth}{!}{\includegraphics{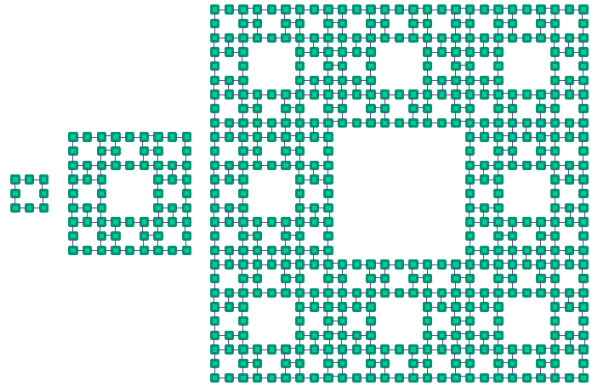}}
  \caption{(Color online) First few generations of a Sierpinski carpet with dimension $\log[8]/\log[3] \approx 1.9$: one starts from a square of $8$ qubits, a $3\times 3$ lattice with the middle qubit removed; then each site is replaced by the original, first-generation square; and so on. Any dimension between 1 and 2 can be obtained by adjusting the total size and missing fraction in the first-generation square: if the length of the square is $n$ qubits and the length of the missing core is $l$ qubits, the effective dimension is $\log[n^2-l^2]/\log[n]$. Starting from a three-dimensional cube, one can obtain any dimension between 2 and 3, and so on.} \label{fig_3}
\end{figure}

One way to obtain noninteger dimensions, without losing local
structure, is to use fractal geometries. Indeed, the proposed SQ
architectures can be used to implement the Ising model on a
well-studied fractal that can be used to probe noninteger
dimensions between $d=1$ and $d=2$, namely, the \emph{Sierpinski
carpet} \cite{Fractals_book}. The relevant construction is
illustrated and explained in Fig.~\ref{fig_3}. Note that a
two-dimensional lattice of qubits without any qubits missing can
be used to generate any desired dimension between 1 and 2. This is
achieved by decoupling any unneeded qubits from the rest of the
lattice using the tunable couplers, effectively removing these
qubits from the system and creating holes in their place.
Alternatively, the Sierpinski carpet can be generated by
engineering the connections between the different qubits on the
chip, even if the qubits are not arranged in a two-dimensional
lattice (as explained in Fig.~2).

\section{Exotic Topologies}

A substantial amount of theoretical work has been devoted to the
critical behavior of spin lattices with nontrivial topology
\cite{Ising_ExoticTopo}. Experimental realizations have been
possible with thin-film materials
\cite{Mobius_molecule,Topo_review}, but with limited
controllability of subsystems. An open question in this area is
whether there is any general principle connecting the properties
of a many-body system with the curvature or the topology of the
surface on which the system resides. It would therefore be
desirable to have small-scale lattices with local addressability,
whose underlying topology is also controllable.

With the designs proposed here, this experimental goal becomes
more feasible. For example, it would be possible to engineer a
spin system whose topology is either a \emph{torus} or a
\emph{Klein bottle}. These two lattice topologies can be
engineered as follows: we start from, say, the $3 \times 3$
lattice shown in Fig.~\ref{fig_2}(a). In addition to the
connections shown, we impose periodic boundary conditions along
the vertical direction, i.e.~we connect qubits (1-7), (2-8) and
(3-9). The type of connections made along the horizontal direction
then differentiates between a torus and a Klein bottle. For the
torus we connect qubits (1-3), (4-6) and (7-9); while for the
Klein bottle we connect (1-9), (4-6) and (3-7). Adding the
connections required to implement either the torus or the
Klein-bottle topology (or even both sets of connections together)
requires the addition of only one layer of couplers, in principle.

The simplest model one could study on these structures is the
Ising model, where nearest neighbors $\langle i,j \rangle$
interact via a $\hat{\sigma}_i^z \hat{\sigma}_j^z$ coupling term.
One could also apply an effective transverse field, i.e.~a
$\hat{\sigma}_i^x$ term in the Hamiltonian, of strength $\lambda$,
on every qubit and obtain signatures of a \emph{quantum} phase
transition \cite{Sachdev} as the parameter $\lambda$ is varied.
One of the most telling signs of the system passing through a
`critical' value $\lambda_c$, is that the \emph{entanglement}
properties of the ground state wave function change quite
drastically \cite{Fazio_review}. Even small systems of, say,
$N=4,9,16$ qubits, will display cusps of increasing sharpness in
some entanglement measure, such as the concurrence between two
neighboring qubits. Such entanglement properties can be measured
relatively easily in the proposed architecture. Since the coupling
strengths are all tunable, the qubits can all be decoupled from
one another. Then, by selectively turning on and off certain
coupling strengths and performing the required quantum gates, the
necessary multi-qubit observables can be measured, e.g., following
quantum state tomography or entanglement witness measurement
protocols. Alternatively, if one could simultaneously couple all
the qubits to a common resonator, one could follow the ideas
proposed in Ref.~\cite{Wang} to detect the quantum  phase
transition through the response of the resonator to an external
probe.

\section{Bose-Hubbard Models}

The architectures described above for spin lattice systems in
nonconventional dimensions and topologies can also be implemented
using resonators, such that the result is a Bose-Hubbard system.
In such systems, one can study the transport properties of the
excitations, which play the role of bosonic particles. In
particular, one can investigate the superfluid--Mott-insulator
phase transition and the Anderson-localization phase transition.

In order to investigate the superfluid--Mott-insulator phase
transition, one needs inter-particle interactions. Such
interactions can be engineered by coupling each resonator to a
qubit, similarly to the proposals of Ref.~\cite{Hartmann}. In
order to investigate the Anderson-localization phase transition,
one needs disorder in the single-site energies or nearest-neighbor
coupling strengths (i.e.~hopping matrix elements). Engineering
such disorder is straightforward with superconducting circuits,
where each site energy and each coupling strength is individually
tunable. A recent theoretical work also proposed the possibility
of engineered time-reversal symmetry breaking in a system of
coupled superconducting resonators \cite{Koch}. This idea can also
be implemented in the systems of arbitrary dimension or topology
proposed in our work.

An important quantity in the study of many-body physics is the
two-point correlation function of the form $\langle a^{\dagger}_i
a_j \rangle$. The tunability of the coupling strengths in SQ
systems enables one to measure this quantity relatively
straightforwardly. One starts by turning all the couplings off. By
measuring the number of excitations in each resonator, and
repeating the experiment a large number of times, one obtains
$\langle a^{\dagger}_i a_i \rangle$. If then one couples only
resonators $i$ and $j$ with coupling strength $J$ as described by
$\hat{H}_2$, the number of excitations as a function of time is
given by
\begin{widetext}
\begin{equation}
\langle a^{\dagger}_i(t) a_i(t) \rangle = \langle a^{\dagger}_i(0)
a_i(0) \rangle \cos^2\left(\frac{Jt}{\hbar}\right) + \langle
a^{\dagger}_j(0) a_j(0) \rangle
\sin^2\left(\frac{Jt}{\hbar}\right) - \frac{i}{2} \langle
a^{\dagger}_i(0) a_j(0) - a^{\dagger}_j(0) a_i(0) \rangle
\sin\left(\frac{2Jt}{\hbar}\right).
\end{equation}
\end{widetext}
One can therefore use the resulting oscillations in order to
extract the imaginary part of $\langle a^{\dagger}_i(0) a_j(0)
\rangle$. Repeating the same procedure with a $\pi/2$ phase shift
applied to one of the resonators, allows the extraction of the
real part of $\langle a^{\dagger}_i(0) a_j(0) \rangle$.

\section{Quench Dynamics and Topology-Quench Experiments}

The \emph{in-situ} tunability of the parameters in SQ circuits
enables one to perform quench-related experiments, where the
parameters are changed from some initial configuration to a
different one. What starts out being the ground state or
thermal-equilibrium state then becomes an excited state that tends
to relax to the new ground state or thermal-equilibrium state. How
this relaxation takes place, and whether it is possible at all
under the constraints imposed by conservation laws has been a
subject of extensive studies in the past
\cite{Fazio_review,Dziarmaga}. The quintessential example of such
quench problems is the Kibble-Zurek mechanism \cite{KibbleZurek},
which describes defect formation in systems that are quenched from
one thermodynamic phase into another and is proposed as the
mechanism for pattern formation, such as galaxy creation, in the
early universe. Using the SQ architecture discussed here, one can
quench the parameters across any of the phase transitions
mentioned above and analyze the resulting dynamics of the system.

We also propose the idea of implementing \emph{topology-quench}
experiments, in which a system is initially embedded in one
topology and, subsequently, its internal interactions are changed
in order to obtain a different topology. The idea can be explained
using the $3 \times 3$ lattice of Fig. \ref{fig_2}(b), which can
be turned into a torus or a Klein bottle, as explained above. The
two different topologies differ by two of the connections
implementing the boundary conditions: (1-3) and (7-9) versus (1-9)
and (3-7). A topology quench can be performed by switching off the
torus-generating connections and then switching on the Klein
bottle-generating connections. It is known that the partition
functions of the Ising model in these two different topologies are
not the same \cite{Ising_ExoticTopo}, and we would therefore
expect to see signatures of the different orders as we change from
one topology to the other. These may be manifested by the
magnetization and thermal entanglement properties
\cite{Fazio_review}. In this case, it is the actual topology of
the underlying lattice structure that changes as one set of
interactions is turned off and another is turned on. The
reordering of the quantum state via the Kibble-Zurek mechanism
will consequently have to follow this change of topology of the
lattice.

\section{Typical experimental parameters}

Superconducting flux qubits typically have minimum gaps [i.e.~the
parameter $\Delta$ in Eq.~(1)] in the range 3-7 GHz \cite{You}.
Recently, new qubit designs have made this parameter tunable
\emph{in situ}, with a minimum value of essentially zero
\cite{Paauw}. The parameter $\Delta$ can therefore be tuned to any
value between 0 and 7 GHz. By changing the externally applied flux
in the large qubit loop, the parameter $\epsilon$ can be tuned to
any value between 0 and values higher than 20 GHz.

The coupling strength between a qubit and a coupler can be
designed through the kinetic inductance arising from shared
superconducting segments or Josephson junctions. Coupling
strengths close to 1 GHz have been achieved \cite{Niskanen}, and
with straightforward parameter changes it should be possible to
reach values of about 2 GHz. Couplers typically have minimum gaps
$\Delta$ in the range 10-20 GHz. These parameters can result in a
maximum inter-qubit coupling strength on the order of a few
hundred MHz \cite{Averin} (Note that, by the design of the circuit
under consideration, the effective inter-qubit coupling strength
is tunable). It is therefore possible to explore parameter
combinations extending from the regime where the single-spin
energies (including both $\sigma_x$ and $\sigma_z$ components) are
larger than the interaction energy to the regime where the
opposite is true. This is the region where phase transitions are
expected to occur, suggesting that the experimental investigation
of these phase transitions and related critical phenomena in
superconducting qubit circuits is feasible.

Superconducting resonators with frequencies in the few-Gigahertz
range, which is the natural range to use in qubit circuits, can be
designed and fabricated with high controllability. In
qubit-resonator circuits, coupling strengths in the range of
10-100 MHz are common, and recent experiments have achieved
coupling strengths close to 1 GHz \cite{FornDiaz}. Since
superconducting couplers have similar structure to qubits, similar
resonator-coupler coupling strengths can be expected. The above
numbers indicate that inter-resonator coupling strengths
(i.e.~hopping coefficients) and on-site inter-particle interaction
coefficients on the order of a few hundred MHz should be
achievable. Since the competition between inter-site hopping and
on-site interactions governs the Superfluid--Mott-Insulator
transition, one should be able to access the
Superfluid--Mott-Insulator transition in the proposed
architecture. Disorder of magnitudes smaller than or larger than
hundreds of MHz can also be achieved, implying that the
Anderson-localization transition can also be investigated.

Typical decay rates for both qubits and resonators are on the
order of 1 microsecond in superconducting circuits. It has been a
challenge to experimentally fabricate multiple qubits on one chip
where all of the qubits have coherence times at that scale.
However, it is expected that in the future long coherence in
multi-qubit circuits will be possible. When that goal is achieved,
it will mean that the overall coherence timescale will be long
compared to typical parameters in the Hamiltonian, which are on
the order of tens of nanoseconds or shorter.

\section{Conclusions and Outlook}

Facilitated by high levels of controllability and steadily
improving coherence properties, superconducting qubits and
resonators are finding various potential applications in quantum
information processing and condensed-matter physics. For example,
there have recently been a number of proposals for using them as
quantum simulators \cite{Buluta}. In this context, the single
qubit/resonator controllability and readout is a key advantage of
superconducting qubits compared to microscopic simulators, such as
natural atoms or ions.

In this work, we have added to the list of feasible potential
applications of superconducting circuits the engineering of
lattice systems in arbitrary dimension and topology. In
particular, we have proposed to engineer spin lattices in integer
dimensions $d\ge 3$, fractal dimensions and nonconventional
topologies. We have also discussed how Bose-Hubbard lattice
systems in similar exotic dimensions and topologies can be
implemented using superconducting resonators. An advantage of SQ
systems in this context, which is particularly exploited in our
proposals, is the high level of connectivity \cite{Our_work}
between superconducting qubits, resonators and hybrid
qubit-resonator systems. Furthermore, the \emph{in-situ}
tunability of the parameters allows for the design of quench, or
even \emph{topology quench}, experiments: in such experiments the
internal reordering of the system could be observed as the
connections between lattice sites are changed. In this case, the
very topology of the underlying lattice could be quenched, thereby
opening the way to a rather different type of quantum quench
experiments.

We should emphasize that our proposal for investigating signatures
of phase transitions in the Ising model does \emph{not} require
multi-qubit quantum coherence. Additionally, fluctuations in the
parameters can be tolerated for purposes of analyzing the presence
or absence of phase transitions. Such experiments should therefore
be easier to realize in the near future. Experiments combining
both scalability and long multi-qubit coherence times are expected
in the coming few years, at which point the investigation of
quantum phase transitions and critical phenomena using
superconducting lattice systems can also be realized.

We would like to thank M. Blencowe and J. R. Johansson for useful
discussions. This work was supported in part by the EPSRC-GB grant
No. EP/G045771/1, DARPA, LPS, NSA, ARO, NSF grant No. 0726909,
Grant-in-Aid for Scientific Research (S), MEXT Kakenhi on Quantum
Cybernetics, and Funding Program for Innovative R\&D on S\&T
(FIRST).


\end{document}